# Gerchberg-Saxton Optimized Silicon Metasurface for Sinusoidal Structured-Light Three-Dimensional Imaging

Gengyuan Ye [1] and Shuming Wang [2,*]

[1] *Nanjing Jinling High School, Zhongshan Road 169, Nanjing 210005, China*

[2] *School of Physics, Nanjing University, Nanjing 210093, China*

**Abstract:** Three-dimensional (3D) optical imaging based on structured illumination has attracted considerable attention in applications including machine vision, robotics, and computational sensing. However, conventional structured-light systems generally rely on bulky projection components, such as digital micromirror devices and diffractive optical elements, which limit system miniaturization and integration. Here, we demonstrate a compact structured-light 3D imaging platform enabled by a silicon dielectric metasurface operating at a near-infrared wavelength of 940 nm. The desired sinusoidal fringe illumination is encoded into the metasurface through a Gerchberg-Saxton (GS) phase optimization algorithm, which converts the target far-field distribution into a nanoscale phase profile and subsequently into silicon nanopillar geometries. The fabricated metasurface generates high-quality sinusoidal structured illumination, which is integrated with a four-step phase-shifting reconstruction method to recover 3D object profiles. The proposed approach provides a compact and scalable solution for miniaturized structured-light projection and optical 3D sensing.

[*]Correspondence and requests for materials should be addressed to S. M. Wang (wangshuming@nju.edu.cn)

## I. INTRODUCTION

Three-dimensional (3D) imaging technologies have become essential for a wide range of applications, including industrial inspection, augmented reality, and biomedical sensing. Among various depth sensing techniques, structured-light imaging has emerged as one of the most accurate approaches due to its high spatial resolution and robustness against environmental variations. By projecting controlled illumination patterns onto an object surface and analyzing the resulting deformation, precise depth information can be extracted through optical phase retrieval. Despite these advantages, conventional structured-light systems usually require bulky optical components, such as digital micromirror devices (DMDs), liquid crystal spatial light modulators (SLMs), or diffractive optical elements (DOEs), to generate structured illumination [1]. These components increase system complexity, optical alignment requirements, and overall footprint, limiting their implementation in compact and integrated imaging platforms.

Recently, optical metasurfaces composed of subwavelength nanostructures have provided a promising route for manipulating light at the nanoscale. By precisely engineering the spatial phase response of individual meta-atoms, metasurfaces have demonstrated capabilities including beam shaping, holography, and wavefront modulation within ultrathin optical architectures. In particular, dielectric metasurfaces based on high-index materials such as silicon offer low optical loss and efficient phase control in the near-infrared regime, making them attractive candidates for replacing conventional structured-light projection modules [2-6].

In this work, we propose a silicon metasurface-based structured-light system for compact 3D imaging. A Gerchberg-Saxton (GS) optimization algorithm [7] is employed to calculate the required phase distribution for generating sinusoidal fringe patterns at a wavelength of 940 nm. The optimized phase profile is encoded into silicon nanopillar structures through a phase-engineering approach. The generated structured illumination is numerically integrating with a four-step phase-shifting reconstruction method to recover 3D object morphology. This work presents a compact optical platform that combines metasurface-enabled light-field engineering with computational 3D reconstruction.

## II. DESIGN OF GS-OPTIMIZED SILICON METASURFACE

### A. Principle of structured-light generation

The generation of sinusoidal structured illumination requires precise control over the spatial distribution of the optical field. In conventional systems, this functionality is typically achieved using macroscopic projection devices. Here, we replace these components with a silicon dielectric metasurface that directly encodes the desired illumination pattern through nanoscale phase modulation.

The target far-field intensity distribution is defined as [8]

$$I_t(x,y) = I_0 W(x,y)[0.5 + 0.5\cos(2\pi fx + \phi_0)] \tag{1}$$

where $I_0$ is the background intensity, and $f$ and $\phi_0$ denote the spatial frequency and initial phase offset of the sinusoidal pattern, respectively. $W(x,y)$ is an aperture function that defines the spatial extent of the generated fringe pattern, with $W(x,y) = 1$ inside the desired rectangular region and $W(x,y) = 0$ outside this region.

To obtain the required metasurface phase distribution, a GS iterative algorithm is employed. During each iteration, the optical field is alternatively propagated between the metasurface plane and the far-field plane while imposing constraints on the phase and amplitude distributions. The optimization process can be expressed as:

$$U_n(x,y) = A_n(x,y)e^{i\Phi_n(x,y)} \tag{2}$$

where $\Phi_n(x,y)$ represents the retrieved phase distribution after the $n$th iteration. After convergence, the optimized phase map is converted into physical silicon nanopillar geometries according to the simulated phase response of individual meta-atoms.

## B. Metasurface unit-cell design

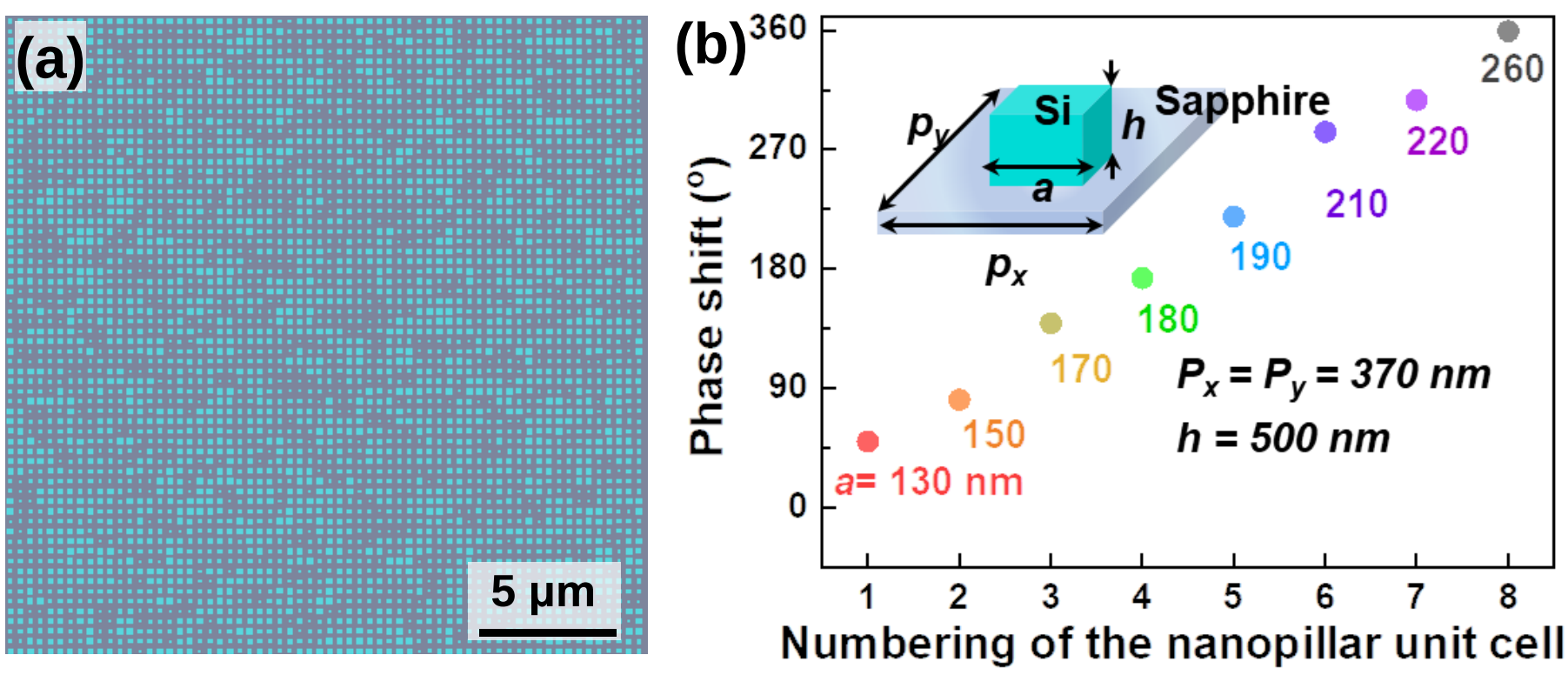

Figure 1. (a) Partial layout of the designed silicon metasurface. (b) Silicon meta-atoms with varying sizes for achieving a full 0-2$\pi$ phase shift. The inset shows the geometry of a single metasurface unit-cell.

The metasurface consists of an array of rectangular silicon nanopillars with a lattice period of $P_x$ = $P_y$ = 370 nm and a height of $h$ = 500 nm on the sapphire substrate (Figure 1(a)). The lateral dimensions of the nanopillars are varied from 130 nm to 260 nm, providing a full 2π phase modulation range at the operating wavelength of 940 nm (Figure 1(b)). The relationship between nanopillar geometry and phase response is obtained using finite-difference time-domain (FDTD) simulations. By selecting appropriate nanopillar dimensions, the desired phase distribution generated by the GS algorithm is accurately encoded into the metasurface.

## III. METASURFACE-GENERATED SINUSOIDAL STRUCTURED ILLUMINATION

In the design workflow of the proposed metasurface structured-light generator, the target sinusoidal fringe pattern is first defined as shown in Figure 2(a). The GS algorithm is subsequently employed to retrieve the required phase distribution (Figure 2(b)). The optimized phase map is then encoded into a silicon metasurface consisting of a 320×180 array of rectangular nanopillars. The designed metasurface contains 57600 silicon nanostructures arranged over an area of 118.4 μm×66.6 μm (refer to Figure 1(a)), corresponding to a lattice period of 370 nm in both the *x* and *y* directions. The optical response of the designed metasurface is evaluated using FDTD simulations. The calculated far-field intensity distribution exhibits a sinusoidal structured-light pattern (Figure 3(a)) that agrees well with the predefined target distribution generated by the GS optimization algorithm (Figure 2(a)). The generated fringe pattern has a period of 16 μm, resulting in approximately 80 fringe periods within the simulated field of view of 1.28 mm×0.72 mm along the *x* and *y* directions, respectively.

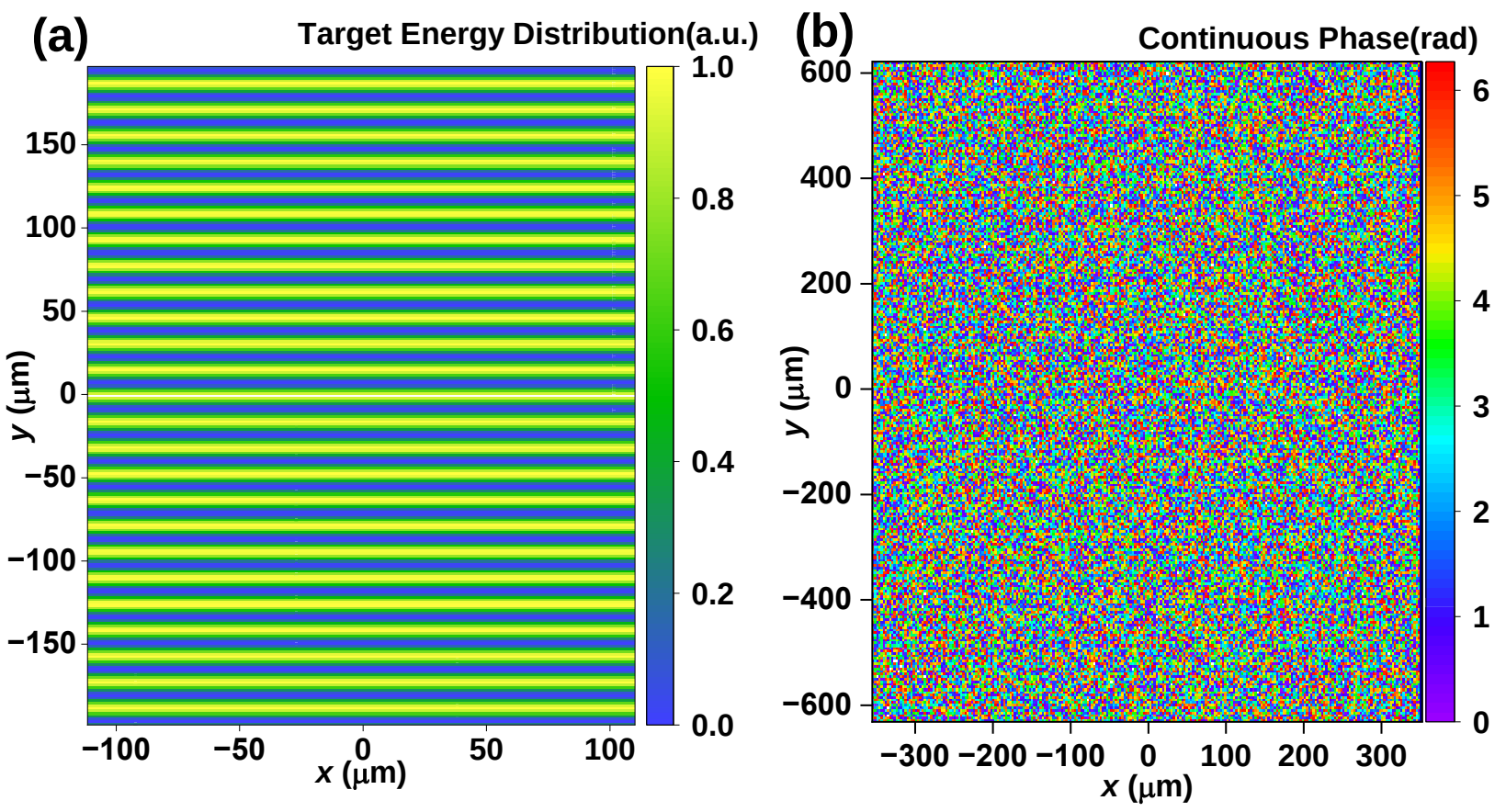

Figure 2. (a) Target far-field intensity distribution of the sinusoidal structured-light pattern. (b) Corresponding target phase distribution obtained from the GS optimization algorithm.

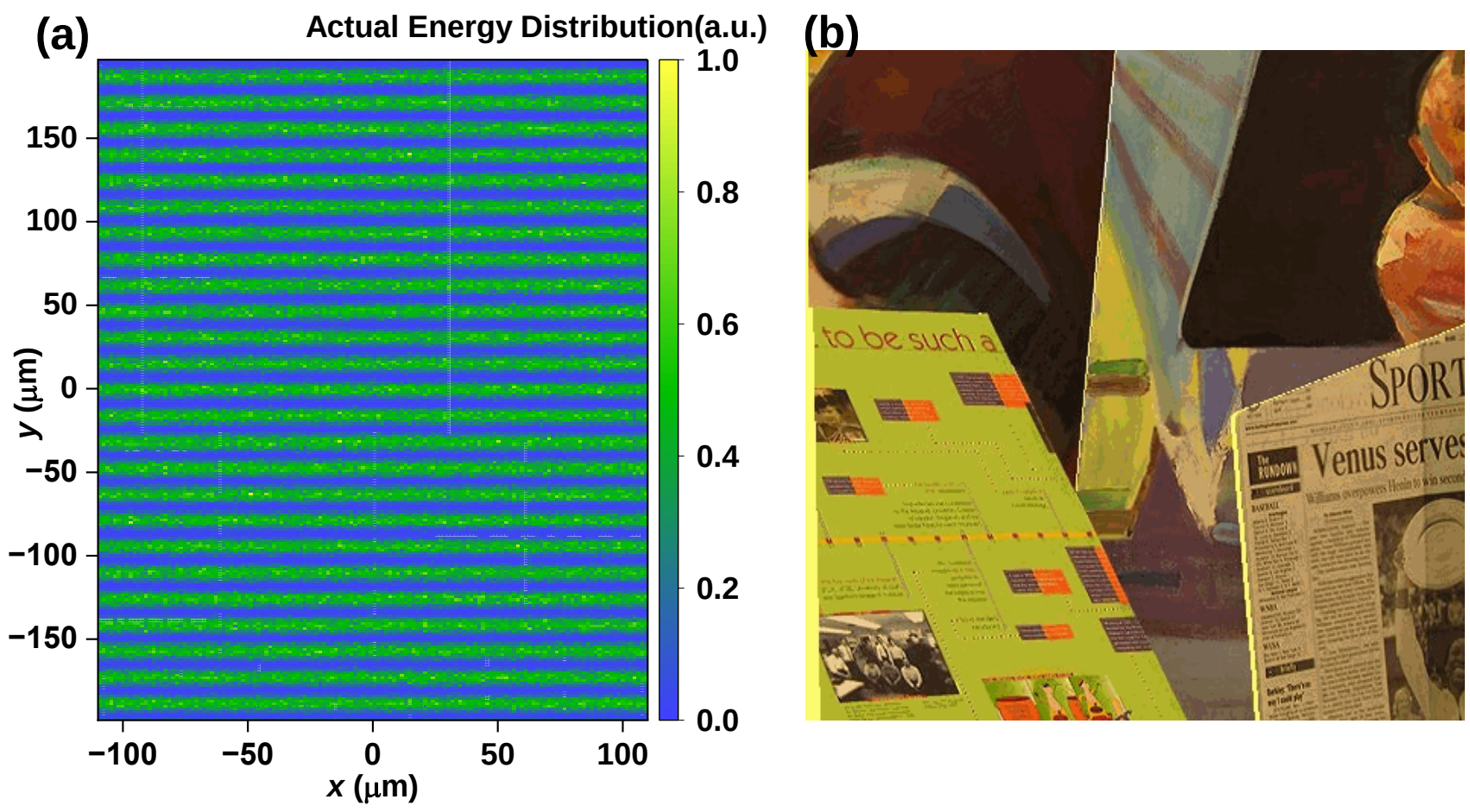

Figure 3. (a) FDTD-calculated far-field intensity distribution generated by the designed metasurface. (b) Ground-truth 3D object used for structured-light reconstruction.

## IV. FOUR-STEP PHASE-SHIFTING THREE-DIMENSIONAL RECONSTRUCTION

The generated sinusoidal structured light is numerically projected onto the target object, and the resulting fringe deformation caused by surface height variations is calculated. The intensity distribution of the four incident sinusoidal structured light can be described as [8]:

$$I_n(x,y) = A(x,y) + B(x,y)\cos[\phi(x,y) + \delta_n)] \tag{3}$$

where $I_n(x,y)$ represents the intensity distribution of the $n$-th phase-shifted fringe pattern, $A(x,y)$ denotes the background intensity, $B(x,y)$ represents the fringe modulation amplitude, $\phi(x,y)$ is the phase distribution induced by the object surface height, and $\delta_n = (n-1)\pi/2$ is the introduced phase shift for the $n$-th illumination state, where $n$ =1,2,3,4 (Figure 4(a)). The wrapped phase can be calculated as:

$$\phi(x,y) = \tan^{-1}\frac{I_4 - I_2}{I_1 - I_3} \tag{4}$$

After phase unwrapping and system calibration, the phase distribution is converted into depth information.

When the four sinusoidal fringe patterns are sequentially illuminated onto the ground-true 3D object (Figure 3(b)), the calculated structured-light patterns exhibit clear fringe deformation induced by surface height variations (Figure 4(b)). The reconstructed depth map obtained from Eqs. (3) and (4) is shown in Figure 5(b), which exhibits good agreement with the ground-truth depth distribution shown in Figure 5(a).

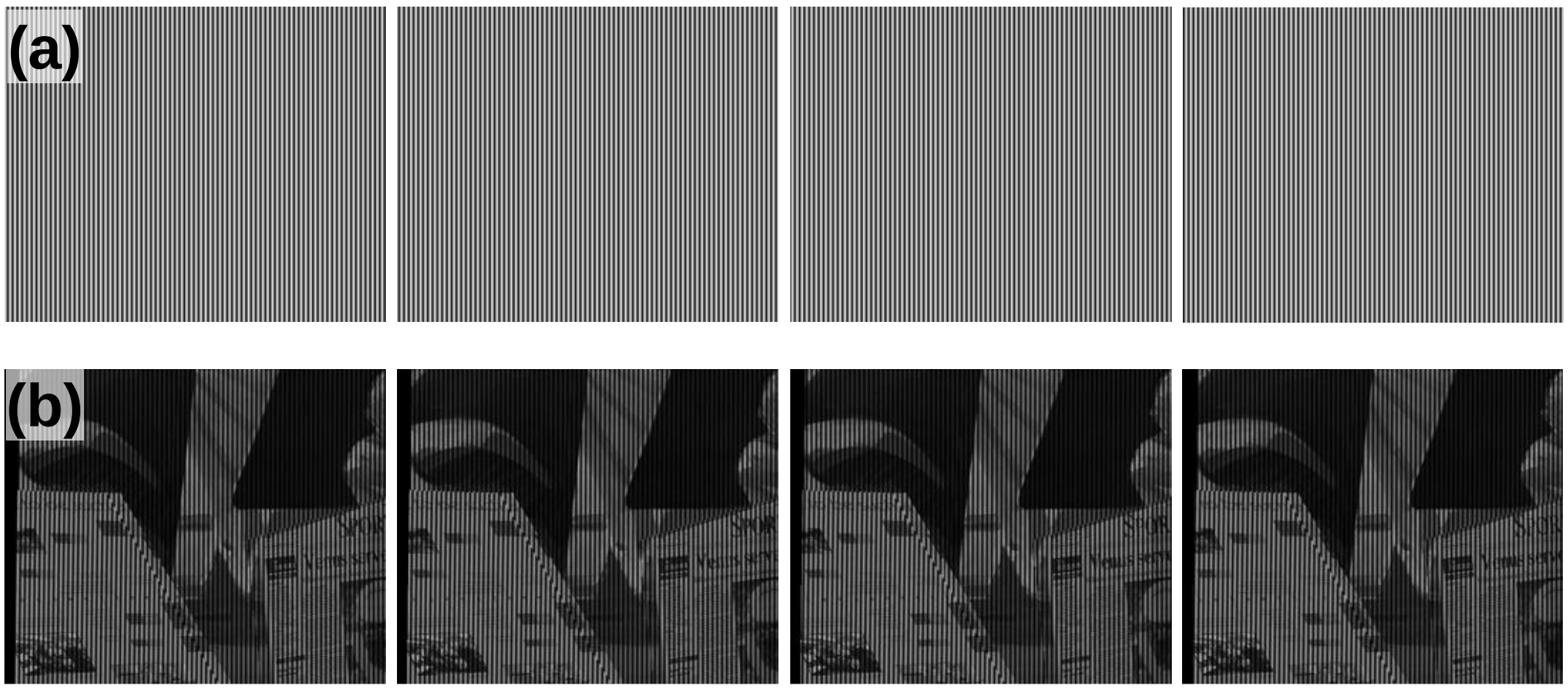

Figure 4. Four-step phase-shifting structured-light reconstruction process: (a) Four incident sinusoidal fringe patterns with different phase shifts. (b) Deformed fringe patterns projected onto the 3D object due to surface height variations.

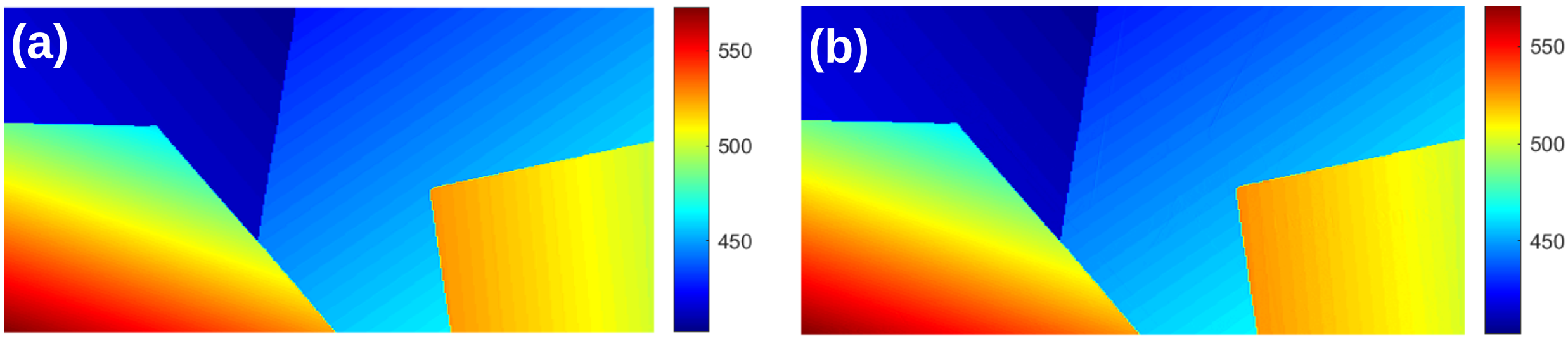

Figure 5. (a) Ground-truth depth map of the 3D object. (b) Reconstructed depth map obtained using the four-step phase-shifting method.

The depth-reconstruction root-mean-square error (RMSE) is quantified using:

$$RMSE = \sqrt{\frac{1}{N}\sum_{i}(z_i - z_i^{\mathrm{GT}})^2} \tag{5}$$

where $z_i$ and $z_i^{\mathrm{GT}}$ represent reconstructed and ground-truth depth values. Under the ideal noise-free condition, the reconstructed depth map achieves an RMSE of approximately 0.14 mm, indicating good agreement with the ground-truth depth distribution. To further evaluate the robustness of the reconstruction method, additive Gaussian noise with standard deviations of $\sigma$ = 0, 0.002, 0.005, 0.01, 0.02, and 0.03 is introduced into the simulated fringe patterns. For a fixed fringe number of 80, the RMSE increases from 0.14 mm to 5.73 mm as the noise level increases from $\sigma$ = 0 to 0.03. In particular, the RMSE remains below 2.34 mm for $\sigma \leq 0.01$, indicating that the proposed phase-shifting reconstruction maintains relatively good accuracy under low-to-moderate noise levels. These results demonstrate the feasibility and noise tolerance of the proposed metasurface-based structured-light 3D imaging approach.

## V. CONCLUSION

In conclusion, we have proposed a compact structured-light three-dimensional imaging framework based on a silicon metasurface. The desired sinusoidal fringe illumination is generated through GS phase optimization and encoded into nanoscale silicon structures operating at a wavelength of 940 nm. The designed metasurface is numerically validated using FDTD simulations, and the generated structured illumination is further employed in a four-step phase-shifting reconstruction framework to retrieve three-dimensional object profiles. By replacing conventional projection optics with an ultrathin metasurface architecture, this approach provides a promising pathway toward compact and integrated optical sensing technologies.